\documentclass[journal=jpccck,manuscript=article]{achemso}
\usepackage[version=3]{mhchem} 
\usepackage{graphicx}
\usepackage{epstopdf}
\usepackage{multirow}
\usepackage{dcolumn}
\usepackage{bm}
\usepackage{subfigure}
\usepackage{color}
\UseRawInputEncoding
\usepackage{amsmath}
\usepackage{amssymb}
\usepackage{amsfonts,color}
\usepackage{amsthm}
\usepackage{xspace}
\usepackage{mciteplus}
\mciteErrorOnUnknownfalse
\usepackage{todonotes}

\usepackage[super]{natbib}
\setcitestyle{super}
\newcommand*{\citen}[1]{%
  \begingroup
    \romannumeral-`\x 
    \setcitestyle{numbers}%
    \cite{#1}%
  \endgroup   
}

\author{N. Yedukondalu}
\email{nykondalu@gmail.com}
\affiliation{Department of Geosciences, Stony Brook University, Stony Brook, New York 11794-2100, USA}
\alsoaffiliation{Joint Photon Sciences Institute, Stony Brook University, Stony Brook, New York 11790-2100, USA}
\author{Aamir Shafique} 
\affiliation{Applied Physics Program, Physical Science and Engineering Division (PSE), King Abdullah University of Science and Technology (KAUST), Thuwal 23955-6900, Saudi Arabia}
\author{S. C. Rakesh Roshan}
\affiliation{Rajiv Gandhi University of Knowledge Technologies, Basar, Telangana-504107, India}
\alsoaffiliation{Department of Physics,  National Institute of Technology-Warangal, Telangana, India} 
\author{Mohamed Barhoumi} 
\affiliation{Laboratoire de la Mati\'ere Condens\'ee et des Nanosciences (LMCN), Universit\'e de Monastir, D\'epartement de Physique, Facult\'e des Sciences de Monastir, Avenue de le \'environnement, 5019 Monastir, Tunisia}
\author{Rajmohan Muthaiah}
\affiliation{School of Aerospace and Mechanical Engineering, University of Oklahoma, Norman, OK 73019, USA}
\author{Lars Ehm}
\affiliation{Department of Geosciences, Stony Brook University, Stony Brook, New York 11794-2100, USA}
\alsoaffiliation{Joint Photon Sciences Institute, Stony Brook University, Stony Brook, New York 11790-2100, USA}
\author{John B. Parise}
\affiliation{Department of Geosciences, Stony Brook University, Stony Brook, New York 11794-2100, USA}
\alsoaffiliation{Joint Photon Sciences Institute, Stony Brook University, Stony Brook, New York 11790-2100, USA}
\author{Udo Schwingenschl\"ogl}
\email{udo.schwingenschlogl@kaust.edu.sa}
\affiliation{Applied Physics Program, Physical Science and Engineering Division (PSE), King Abdullah University of Science and Technology (KAUST), Thuwal 23955-6900, Saudi Arabia}

\date{\today}
\title[An \textsf{achemso} demo]
{ \vspace{-0.5 in} Lattice Instability and Ultralow Lattice Thermal Conductivity of Layered PbIF}

\begin{document}

\begin{abstract}
Understanding the interplay between various design strategies (for instance, bonding heterogeneity and lone pair induced anharmonicity) to achieve ultralow lattice thermal conductivity ($\kappa_l$) is indispensable for discovering novel functional materials for thermal energy applications. In the present study, we investigate layered PbXF (X = Cl, Br, I), which offers bonding heterogeneity through the layered crystal structure, anharmonicity through the Pb$^{2+}$ $6s^2$ lone pair, and phonon softening through the mass difference between F and Pb/X. The weak inter-layer van der Waals bonding and the strong intra-layer ionic bonding with partial covalent bonding result in a significant bonding heterogeneity and a poor phonon transport in the out-of-plane direction. Large average Gr\"uneisen parameters ($\geq$ 2.5) demonstrate strong anharmonicity. The computed phonon dispersions show flat bands, which suggest short phonon lifetimes, especially for PbIF. Enhanced Born effective charges are due to cross-band-gap hybridization. PbIF shows lattice instability at a small volume expansion of 0.1$\%$. The $\kappa_l$ values obtained by the two channel transport model are 20-50$\%$ higher than those obtained by solving the Boltzmann transport equation. Overall, ultralow $\kappa_l$ values are found at 300 K, especially for PbIF. We propose that the interplay of bonding heterogeneity, lone pair induced anharmonicity, and constituent elements with high mass difference aids the design of low $\kappa_l$ materials for thermal energy applications. 
\end{abstract}
\clearpage

\section{Introduction}
The discovery and design of materials with ultralow lattice thermal conductivity ($\kappa_l$) receive tremendous research interest\cite{Juneja2019-ML} because of potential applications in energy conversion technology. Additionally, (ultra)low $\kappa_l$ materials with wide band gaps have promising applications in thermal barrier coatings \cite{MOTEB202021939,LIU2020702}. Ultralow $\kappa_l$ materials with narrow band gaps are suitable for thermoelectric applications.\cite{Narrow-APP-TE-1,Narrow-TE-2-Guo2020,narrow-TE-3,Narrow-TE-4Wu_2020,narrow-5Osei-Agyemang2019} Several strategies\cite{Manoj2018,Moinak2021} have been proposed to design ultralow $\kappa_l$ materials including resonant bonding,\cite{Lee2014} rattling,\cite{Jana2017, Dutta2021,Bhui2021,Juneja2019-Rattling} cation disorder,\cite{Qiu2014} bonding heterogeneity,\cite{Dutta2019,Pal2019,Mukherjee2020} and lone pair induced anharmonicity.\cite{Skoug2011,Chang2018,Rathore2019} Combining two or more of these in one material suppresses $\kappa_l$ effectively. For instance, a lone pair enhances the anharmonicity, while bonding heterogeneity results in anisotropic $\kappa_l$ with poor phonon transport along the weak bonding (soft) lattice direction. Materials with lone pair cation(s) (In$^+$, Tl$^+$, Sn$^{+2}$, Pb$^{+2}$, Bi$^{+3}$, Sb$^{+3}$, etc.)\cite{Rathore2019,Jana2017,Juneja2019-Rattling,Dutta2019} can exhibit enhanced Born effective charges (BECs) due to cross-band-gap hybridization,\cite{TlXDu2010,InXDu2010} causing large splitting between the longitudinal and transverse optical modes and bringing the lattice close to instability.\cite{TlXDu2010,Tachibana2008,Fu2018,Banik2019}


Recently, wide band gap layered semiconductors receive great attention due to potential applications in electronics and optoelectronics in the green and blue wavelength regions.\cite{Lu2020} In particular, PbClF-type (matlockite) materials attract interest because of their applications in photo-detectors and optoelectronic devices.\cite{Chen2004,Manjunath2020,Barhoumi2020,Ye2020} Barhoumi et al. \cite{Barhoumi2019} have investigated the electronic and vibrational properties of several PbClF-type two-dimensional (2D) materials for possible utilization in optoelectronic devices. Tan et al. \cite{Tan2017} have synthesized 2D lead halides and mixed halofluorides to develop low-cost, high-sensitivity, low-noise, and flexible ultra-violet photodetectors. Understanding the finite temperature lattice dynamics and phonon transport is critically important for deploying PbXF (X = Cl, Br, I) as low $\kappa_l$ materials. However, only few studies have focused on PbClF-type materials such as 2D-SrBrF\cite{TanJing2021} and the bulk crystals CaClF and SrClF.\cite{Lavanya2021} 

PbXF (X = Cl, Br, I) crystallizes in the tetragonal space group $P4/nmm$ with two formula units in the primitive cell.\cite{FLAHAUT1974124} F occupies the Wyckoff position 2a ($3/4$, $1/4$, 0) and Pb and X occupy the Wyckoff position 2c ($1/4$, $1/4$, $v/u$). The Pb$^{2+}$ cations have $6s^2$ lone pairs. The structure consists of X-Pb-F-F-Pb-X layers, with strong intra-layer ionic bonding ($ab$-plane), which are connected in the $c$-direction by weak inter-layer van der Waals (vdW) bonding, similar to MXF (M = Ca, Sr, Ba and X = Cl, Br, I) at ambient conditions.\cite{Yedu2011,kanchana2013} Panels a and b of Figure $\ref{fig:str}$ show the matlockite structure of PbIF. The combination of intra-layer ionic bonding with inter-layer vdW bonding results in bonding heterogeneity and thus in anisotropic phonon transport, directly affecting the phonon group velocities. As there are no studies available that explore the potential of PbXF (X = Cl, Br, I) in optoelectronic devices and thermal energy management, we systematically investigate the electronic structure, anharmonic lattice dynamics, and phonon transport.

\section{Computational details and methodology}
First principles calculations are performed using the Vienna ab initio simulation package (VASP)\cite{Kresse1996}. The electron-electron interaction is treated by the PBEsol exchange-correlation functional within the generalized gradient approximation and the electron-ion interaction by the projector augmented wave pseudopotential approach. The following valence electrons are considered: F 2s$^2$, 2p$^5$,  Cl 3s$^2$, 3p$^5$,  Br 4s$^2$, 4p$^5$,  I 5s$^2$, 5p$^5$, and Pb 5d$^{10}$, 6s$^2$, 6p$^2$. For the structural optimization and computation of the 2$^{nd}$ order elastic constants an energy cutoff of 600 eV is used for the plane wave basis and a spacing of 2$\pi$ $\times$ 0.025 \AA$^{-1}$ for the k-mesh of the Brillouin zone integration. Electronic structure calculations are performed using the Tran-Blaha modified Becke-Johnson (TB-mBJ) \cite{Tran2009} potential as implemented in WIEN2k,\cite{Blaha2002} and the Heyd-Scuseria-Ernzerhof (HSE)\cite{HSE} functional as implemented in VASP.\cite{Kresse1996} 


The finite temperature lattice dynamics and thermal conductivity of PbXF (X = Cl, Br, I) are studied using the temperature dependent effective potential (TDEP) method.\cite{TDEP_PRB_2011} To compute the harmonic (second order) and anharmonic (third order) inter-atomic force constants (IFCs) at 300 K we perform ab initio molecular dynamics (AIMD) simulations using VASP, comprising 5000 steps of 1 fs (5 ps total duration) with the temperature controlled by a Nos\'e-Hoover thermostat. A 4 $\times$ 4 $\times$ 2 (192 atoms) supercell is used and only the $\Gamma$-point is considered in the Brillouin zone integration. For calculating the second and third order IFCs, interactions up to 9$^{th}$ nearest neighbours are included to ensure convergence of the phonon dispersion and lattice thermal conductivity. TDEP includes finite temperature effects on the second and third order IFCs, which are critical for highly anharmonic low $\kappa_l$ materials. The lattice thermal conductivity is calculated by iteratively solving the  Boltzmann transport equation (BTE), including three-phonon and isotope (natural distribution) scattering on a $19\times19\times19$ q-mesh using both TDEP\cite{TDEP_PRB_2011} and ShengBTE\cite{shengbte2014}. In the case of ShengBTE the finite displacement method with a $3\times3\times2$ supercell is employed to calculate the second and third order IFCs by Phonopy \cite{phonopy2015} and the thirdorder.py script\cite{thirdorder}, respectively.

\section{Results and Discussion}
\subsection{Crystal structure, bonding, and elastic properties}
\begin{figure}
\centering
\includegraphics[width=6.0in,height=5.0in]{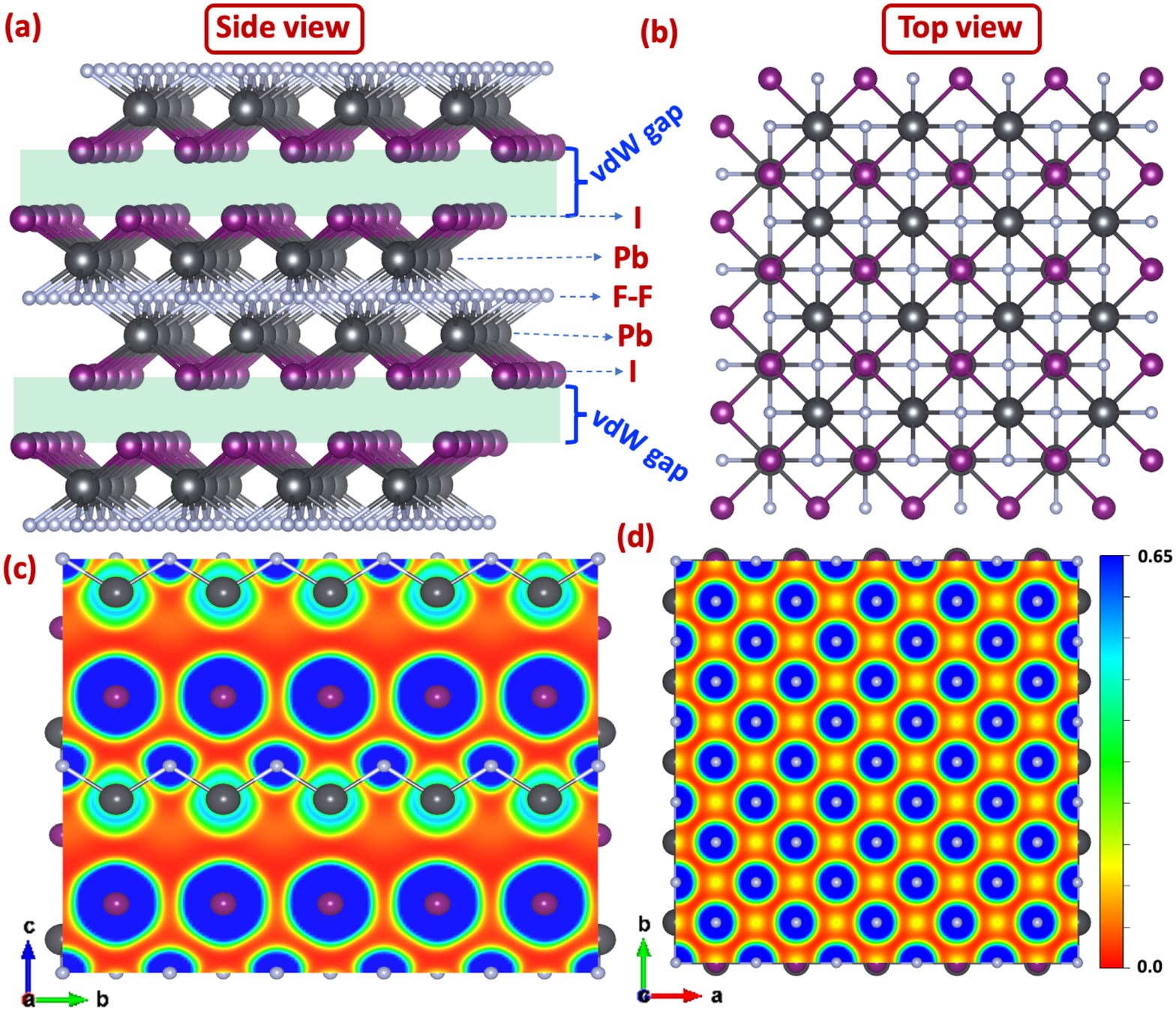}
\caption{Layered crystal structure of PbIF in (a) side and (b) top views. ELF of PbIF in the (c) $bc$- and (d) $ab$-planes.}
\label{fig:str}
\end{figure}

Since the matlockite structure of PbXF (X = Cl, Br, I) is layered (Figure \ref{fig:str}a,b), we optimize it using the PBEsol, DFT-D2, and non-local optB88-vdW and optB86b-vdW methods to capture the vdW interaction. The experimental volume is well reproduced by the PBEsol and optB86b-vdW methods (Supporting Information Tables S1 $\&$ S2). We choose the equilibrium structure obtained by the PBEsol functional to compute the Bader charges, elastic properties, lattice dynamics, and thermal conductivity. The calculated electron localization function (ELF) shows a strong intra-layer ionic bonding with partial covalent bonding along with stereochemical activity of the Pb$^{2+}$ 6s$^2$ lone pair (Figure \ref{fig:str}c,d). To understand the bonding nature quantitatively, we calculate the Bader charges of the Pb, F, and X ions, finding that the two valence electrons of the Pb$^{2+}$ cation are not completely transferred to the F and X anions (Table \ref{ELFT}), confirming a mixed ionic and covalent bonding character. The charge transfer from Pb to Cl (-0.7), Br (-0.6), and I (-0.5) decreases with the size of the anion, indicating that the covalent contribution to the bonding increases, while the bonding between Pb and F is mainly ionic. The bonding character of PbXF (X = Cl, Br, I) thus changes with the size (electronegativity) of the X anion. 



The elastic constants of crystalline solids provide insights into the bonding strengths, mechanical properties, and stability. The elastic constants obtained at the PBEsol equilibrium volume and at the experimental volume, as presented in Table \ref{elastic}, agree with the available experimental results as well as with the results of previous calculations.\cite{SIESKIND199875} Due to the tetragonal space group, the elastic tensor has six independent constants (C$_{11}$, C$_{33}$, C$_{12}$, C$_{13}$, C$_{44}$, and C$_{66}$). The computed values fulfill the Born's stability criteria, C$_{11}$ $\textgreater$ 0, C$_{33}$ $\textgreater$ 0, C$_{44}$ $\textgreater$ 0, C$_{66}$ $\textgreater$ 0, (C$_{11}$-C$_{12}$) $\textgreater$ 0, (C$_{11}$ + C$_{33}$ - 2C$_{13}$) $\textgreater$ 0, and [2(C$_{11}$+C$_{12}$) + C$_{33}$+4C$_{13}$] $\textgreater$ 0, indicating mechanical stability at ambient pressure. As Table \ref{elastic} shows C$_{33}$ $\textless$ C$_{11}$ for all three materials, the lattice is softer in the $c$-direction than in the $a$-direction. This is due to the weak vdW interaction between the X-Pb-F-F-Pb-X layers stacked along the $c$-direction. Moreover, we find C$_{13}$ $\textless$ C$_{12}$ and C$_{44}$ $\textless$ C$_{66}$, demonstrating substantial elastic anisotropy, which also originates from the layered crystal structure. Similar trends have been reported for the ultralow $\kappa_l$ materials BiCuChO (Ch = S, Se, Te), which crystallize in the same space group ($P4/nmm$) and realize a layered ZrSiCuAs-type structure.\cite{Ji2016}  


We calculate the bulk (B) and shear (G) moduli from the elastic constants using the Voigt-Reuss-Hill (VRH) approximation\cite{Yedu2011} and then calculate Young's modulus (E) from the obtained B and G values. The calculated B, G, and E values decrease from PbClF to PbBrF to PbIF, pointing to a weakening of the interatomic interactions for increasing size of the X anion (Table \ref{ployc}). Low values indicate that the materials are easily deformed by mechanical stress. Usually, soft materials with heavy elements show low optical phonon frequencies, which enhances the coupling with the acoustic modes, thus enhancing the phonon scattering and lowering $\kappa_l$. Using the obtained B and G values, the longitudinal ($v_l$) and transverse ($v_t$) sound velocities are calculated as $\rho v_l^2$ = B+$\frac{4G}{3}$ and  $\rho v_t^2$ = G, respectively, where $\rho$ is the mass density. 
The Debye temperature ($\theta$) can be estimated from the average sound velocity ($v_m$) as 
\begin{equation}\label{debye}
\theta = \frac{\hbar}{k_B}\left(6\pi^2n\right)^{\frac{1}{3}} v_m,
\end{equation}
where $\hbar$, $k_B$, and $n$ are the reduced Planck constant, Boltzmann constant, and number of atoms per unit cell, respectively. From the obtained $v_l$ and $v_t$ values, $v_m$ is calculated from the relation $\frac{3}{v_m^3}$ = $\frac{1}{v_l^3}$ + $\frac{2}{v_t^3}$.
The presence of low lying optical phonons results in softening of the acoustic phonons, which generates low sound velocities and therefore low $\theta$ ($\propto$ $v_m$). According to the Slack theory, a low $\theta$ value is an indication of low $\kappa_l$. The calculated $v_l$, $v_t$, $v_m$, and $\theta$ values decrease from PbClF to PbBrF to PbIF and a similar trend thus is expected for $\kappa_l$. The mechanical properties of the investigated materials are summarized in Table \ref{ployc}.  

\subsection{Born effective charges and cross-band-gap hybridization}
Owing to the polar nature of PbXF (X = Cl, Br, I), long-range dipole-dipole interaction produces a splitting between the longitudinal optical (LO) and transverse optical (TO) phonons, known as LO-TO splitting. The calculated BECs, which represent the dynamic electron density associated with the covalent bonding, are presented in Table \ref{born}. With decreasing electronegativity of the halogen anion, F to Cl to Br to I, the ionic character of the Pb-X bond decreases, resulting in enhanced in-plane BECs consistent with previously observed trends in layered $\alpha$-PbO\cite{Biswas2020} and BiCuChO (Ch = S, Se, Te)\cite{Ji2016}, which crystallize in the same ($P4/nmm$) space group. The in-plane BECs are close to double of the formal charges (+2 for Pb and -1 for F, Cl, Br, I) due to the Pb$^{2+}$ 6s$^2$ lone pair. Similarly enhanced BECs have been reported for lone pair containing In, Tl, Pb, and Bi halides,\cite{TlXDu2010,InXDu2010} in which the outermost cation $s$-states are fully occupied and the states at the conduction band minimum are mainly derived from the spatially more extended cation $p$-states. This causes a significant cross-band-gap hybridization between the cation $p$- and halogen $p$-states.\cite{TlXDu2010,InXDu2010} 

The electronic band gaps obtained without and with spin-orbit coupling (SOC) are presented in Table \ref{bandgaps}. While the PBE functional underestimates the band gap for semiconductors and insulators, the values obtained by the HSE functional and TB-mBJ potential agree with each other. The value of 4.8 (4.7) eV calculated without (with) SOC for PbClF slightly underestimates the 5.0 eV obtained  by the computationally expensive GW approach and the 5.2 eV obtained by reflection spectroscopy.\cite{Liu2005} The band gap decreases from PbClF to PbBrF to PbIF due to shifts of the cation $p$-states for increasing size of the X anion.   

\begin{figure}
\centering
\subfigure[]{\includegraphics[width=0.4\textwidth]{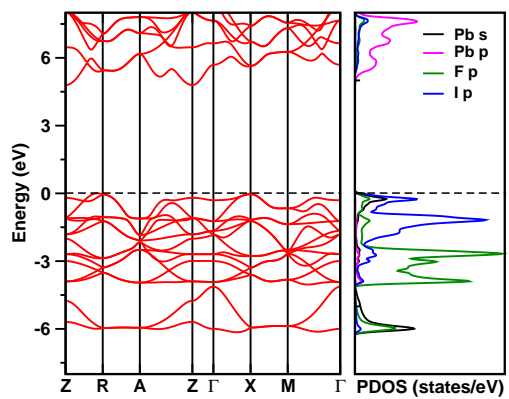}}  \vspace{0.3in} \\ 
\subfigure[]{\includegraphics[width=0.4\textwidth]{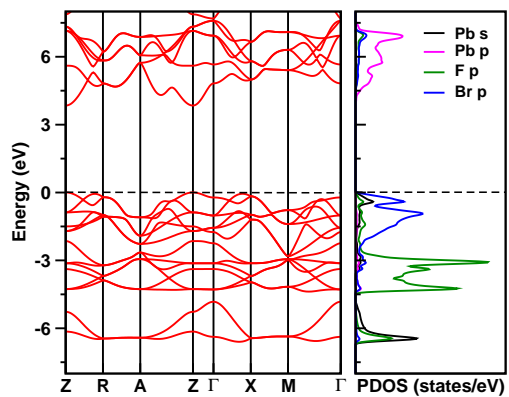}}  \vspace{0.3in} \\
\subfigure[]{\includegraphics[width=0.4\textwidth]{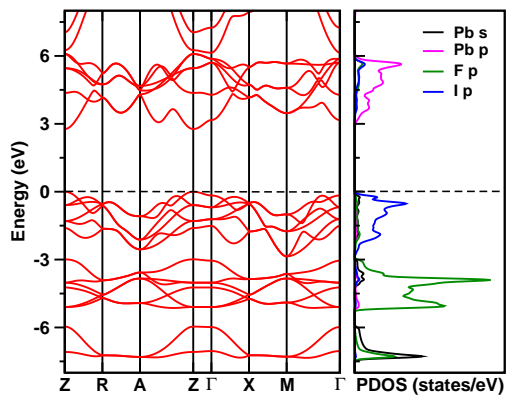}}
\caption{Electronic band structure and projected densities of states of (a) PbClF, (b) PbBrF, and (c) PbIF taking into account the SOC.}
\label{fig:DOS}
\end{figure}

In general, the obtained band structures are similar to those of previous calculations.\cite{Reshak2007,Liu2005,hajHassan2004} As shown in Figure \ref{fig:DOS}, the valence band can be broadly divided into three regions, which are well separated for PbIF in contrast to PbClF and PbBrF. This trend can be attributed to the increasing electronegativity difference between F (3.98, Pauling scale) and Cl (3.16), Br (2.96), and I (2.66). As illustrated in Figure \ref{fig:DOS}, the valence band edge is dominated by the X $p$-states with minor contributions of the Pb $s$-states and F $p$-states, the middle region is dominated by the F $p$-states with small contributions of the X $p$-states, and the lower region arises from strongly hybridized Pb $s$-states and F $p$-states. The dispersive nature of the valence band is the result of hybridization between the fully occupied Pb $s$-states and the X and F $p$-states. The conduction band is dominated by the nominally unoccupied Pb $p$-states, which hybridize with the X and F $p$-states. This gives rise to a considerable cross-band-gap hybridization and, thus, enhancement of the BECs. 

Our results also provide insight into the activity of the Pb$^{2+}$ 6s$^2$ lone pair. When the lone pair is stereochemically (in)active, the projected density of states (PDOS) shows that the occupied cation $s$-states are (narrow) broad and (largely unmixed) well mixed with the anion $p$-states, with intermediate unoccupied cation $p$-states.\cite{Waghmare2003} In the metal chalcogenides AQ (A = Ge, Sn, Pb and Q = S, Se, Te), which contain lone pair cations, the strength of the stereochemical activity decreases for increasing size of the chalcogen atom due to separation between the cation $s$-states and anion $p$-states in the valence band. As shown in Figure \ref{fig:DOS}, the occupied Pb$^{2+}$ $s$-states are narrow and become increasingly separated from the anion (F$^-$ and X$^-$) $p$-states from PbClF to PbBrF to PbIF, developing a pseudo gap. This indicates that the stereochemical activity decreases for increasing size of the X anion, similar to the behavior observed for AQ (A = Ge, Sn, Pb and Q = S, Se, Te).\cite{Waghmare2003}

\subsection{Anharmonic lattice dynamics and thermal conductivity} 
As the phonon frequencies (Raman and IR) at the $\Gamma$-point are vital to understand the lattice dynamics and phonon transport, they have been extensively studied for MXF (M = Ca, Sr, Ba and X = Cl, Br, I).\cite{Lv2017,SUNDARAKANNAN2002,Sorb2013,Yedu_AIP_2015CP,Yedu_IOP_2012CP} These materials possess six atoms in the unit cell, resulting in a total of 18 (3 acoustic and 15 optical) phonon modes. According to group theory, the symmetry decomposition of the $\Gamma$-point phonons for the $P4/nmm$ space group is  $\Gamma_{18}$ = 6E$_g$ $\oplus$ 2A$_{1g}$ $\oplus$ B$_{1g}$ $\oplus$ 6E$_u$ $\oplus$ 3A$_{2u}$, with six Raman active (A$_{1g}$ , B$_{1g}$, E$_g$) and four IR active (A$_{2u}$ , E$_u$) modes, where E$_g$ and E$_u$ are doubly degenerate (Figures S1 $\&$ S2).
\par
\begin{figure}
\centering
\subfigure[]{\includegraphics[width=0.5\textwidth]{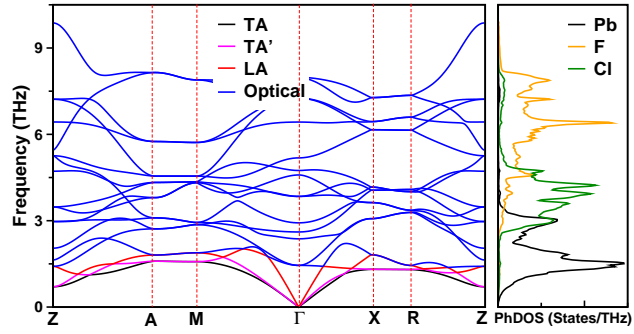}} \vspace{0.45in} \\
\subfigure[]{\includegraphics[width=0.5\textwidth]{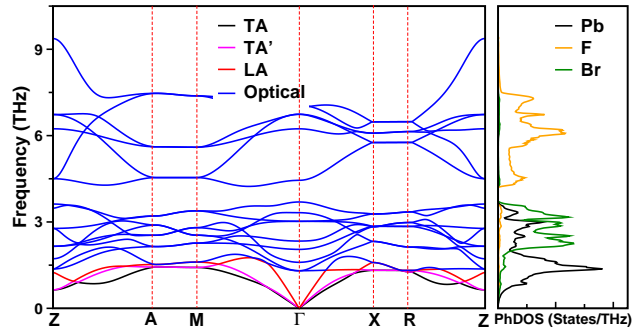}} \vspace{0.45in} \\ 
\subfigure[]{\includegraphics[width=0.5\textwidth]{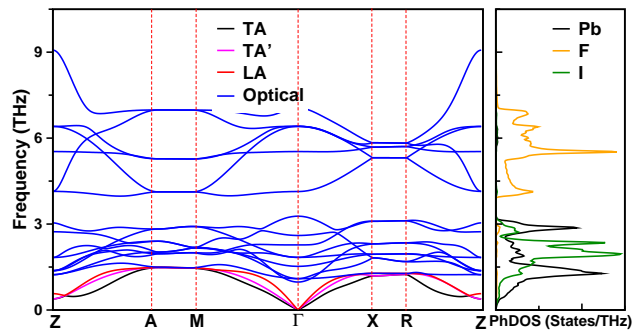}} 
\caption{Phonon dispersion and projected phonon densities of states of (a) PbClF, (b) PbBrF, and (c) PbIF at 300 K.}
\label{fig:Pb-PD}
\end{figure}

 The phonon dispersions and densities of states of PbXF (X = Cl, Br, I) in Figure \ref{fig:Pb-PD} (calculated including the third order IFCs) show no imaginary frequencies, demonstrating dynamical stability at 300 K. The low-(high-)frequency phonons are mainly due to vibrations of the Pb, I, and Br (F and Cl) atoms. The phonon bands are more dispersive for PbClF than PbBrF and PbIF due to the smaller  mass difference between F and Cl as compared to that between F and Br/I, which leads to a phonon band gap in the cases of PbBrF and PbIF as previously reported for MQ (M = Ca, Sr, Ba and Q = S, Se, Te)\cite{Roshan2021}. In the low frequency region, the phonon density of states (Figure \ref{fig:Pb-PD}) is due to overlapping acoustic and low lying optical phonons mainly involving vibrations of the Pb atoms (with contributions of I only in the case of PbIF). The vibrational frequencies of the Cl, Br, and I atoms show a red-shift from PbClF to PbBrF to PbIF due to the increasing mass difference to F. The red-shift is associated with an increasing LO-TO splitting that reduces the dispersions of the phonon bands below 3 THz from PbClF to PbBrF to PbIF (Figure \ref{fig:Pb-PD}). This substantial softening of the acoustic phonons increases the scattering phase space and aids the suppression of $\kappa_l$ (especially for PbIF), similar to the previously observed behavior in layered BiCuSeO.\cite{BiCuSeO2017} 

\begin{figure}
\centering
\subfigure[]{\includegraphics[width=0.5\textwidth]{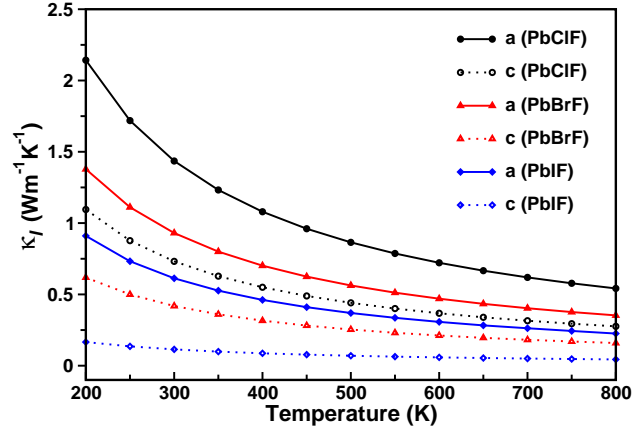}}\vspace{0.3in}
\subfigure[]{\includegraphics[width=0.5\textwidth]{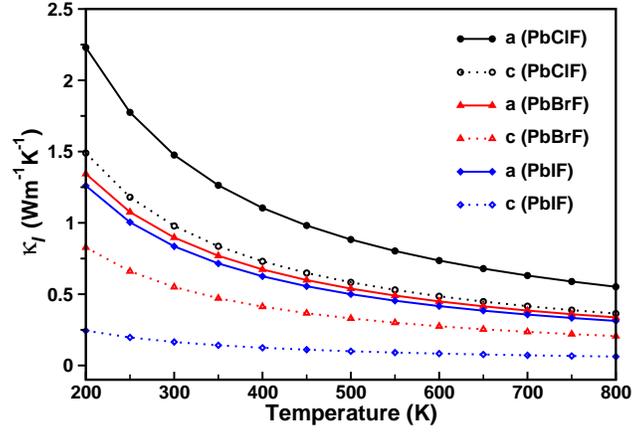}}\vspace{0.3in}
\subfigure[]{\includegraphics[width=0.5\textwidth]{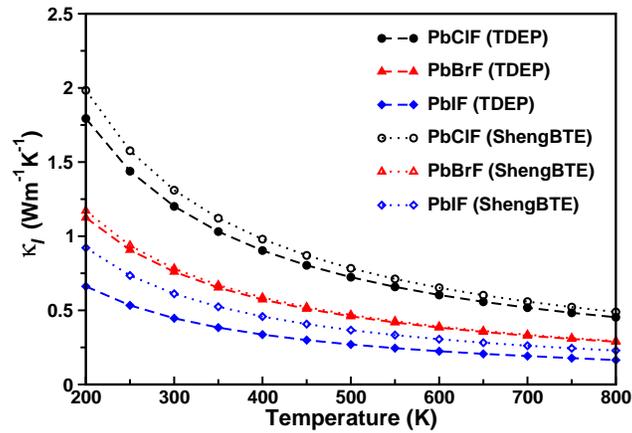}}
\caption{(a, b) Anisotropic and (c) average lattice thermal conductivities as functions of temperature for PbXF (X = Cl, Br, I) using TDEP and ShengBTE by solving the BTE iteratively.}
\label{fig:TC}
\end{figure}

\begin{figure}
\centering
\subfigure[]{\includegraphics[width=0.5\textwidth]{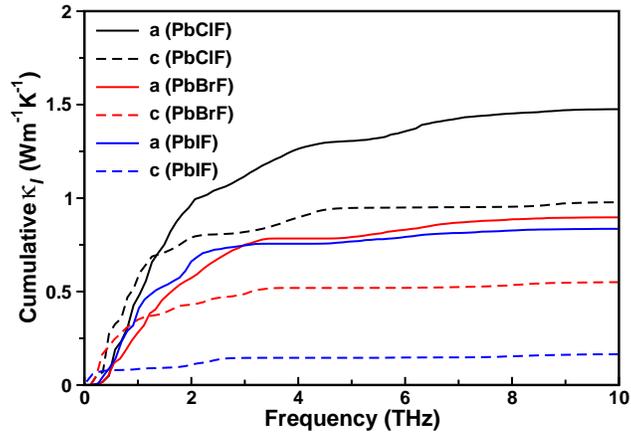}}\vspace{0.3in} \\ \subfigure[]{\includegraphics[width=0.5\textwidth]{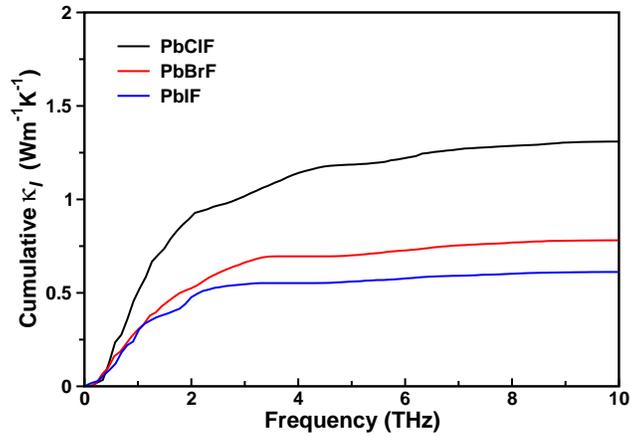}} 
\caption{(a) Cumulative and (b) average cumulative lattice thermal conductivities of PbXF (X = Cl, Br, I)  at 300 K.}
\label{fig:Pb-CK}
\end{figure}

Figure \ref{fig:TC} shows $\kappa_l$ as a function of temperature. Using TDEP (ShengBTE) at 300 K, ultralow $\kappa_l$ values (in Wm$^{-1}$K$^{-1}$) are predicted in the $c$-direction (0.73 (0.98) for PbClF;  0.42 (0.55) for PbBrF; 0.11 (0.16) for PbIF) and significantly higher values in the $a$-direction (1.44 (1.47) for PbClF; 0.93 (0.90) for PbBrF; 0.61 (0.84) for PbIF). This strong anisotropy of $\kappa_l$ originates from the bonding heterogeneity which results in $v_t$ $\textless$ $v_l$ (Table \ref{ployc}). The calculated average $\kappa_l$ values at 300 K are 1.20 (1.31), 0.76 (0.78), and 0.45 (0.61) Wm$^{-1}$K$^{-1}$ for PbClF, PbBrF, and PbIF, respectively. PbXF (X = Cl, Br, I) shows ultralow $\kappa_l$ values in the $c$-direction in the whole temperature range from 200 to 800 K. Accordingly, the $c$-direction gives only minor contributions to the cumulative $\kappa_l$ (Figure \ref{fig:Pb-CK}a). About 70$\%$ of $\kappa_l$ is due to phonons below 2.1 THz for PbClF, 2.2 THz for PbBrF, and 1.85 THz for PbIF (Figure \ref{fig:Pb-CK}b). While the optical phonons contribute significantly to $\kappa_l$, the low dispersions of the acoustic and low lying optical phonon bands result in the ultralow $\kappa_l$ values. 

\begin{figure}
\centering
\subfigure[]{\includegraphics[width=0.5\textwidth]{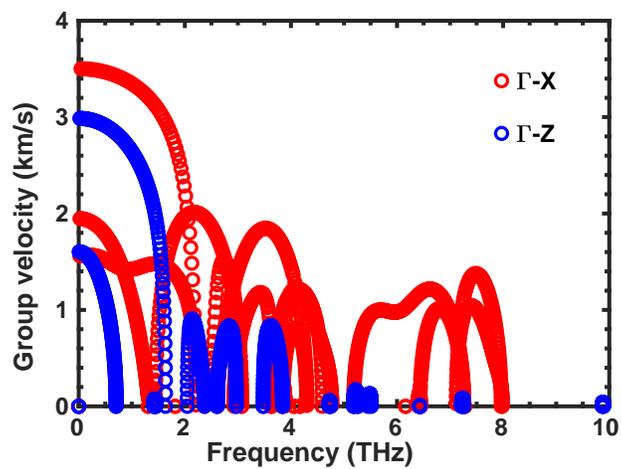}} \\
\subfigure[]{\includegraphics[width=0.5\textwidth]{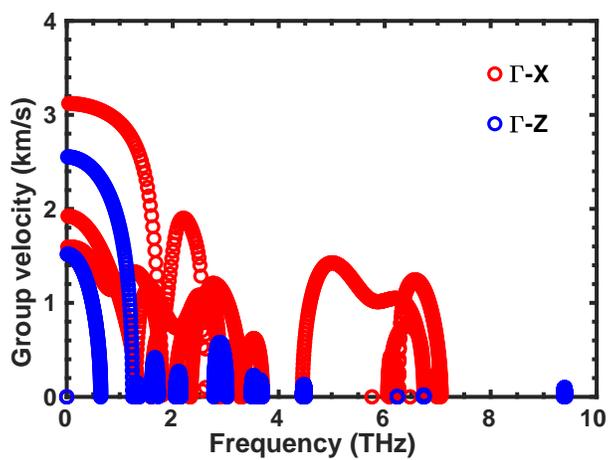}} \\
\subfigure[]{\includegraphics[width=0.5\textwidth]{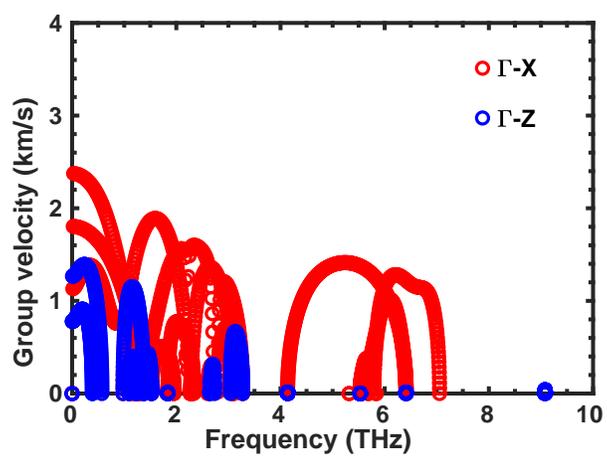}}
\caption{Phonon group velocities of (a) PbClF, (b) PbBrF, and (c) PbIF in the $\Gamma$-X and $\Gamma$-Z directions as functions of frequency.}
\label{fig:GV}
\end{figure}

\begin{figure}
\centering
\subfigure[]{\includegraphics[width=0.5\textwidth]{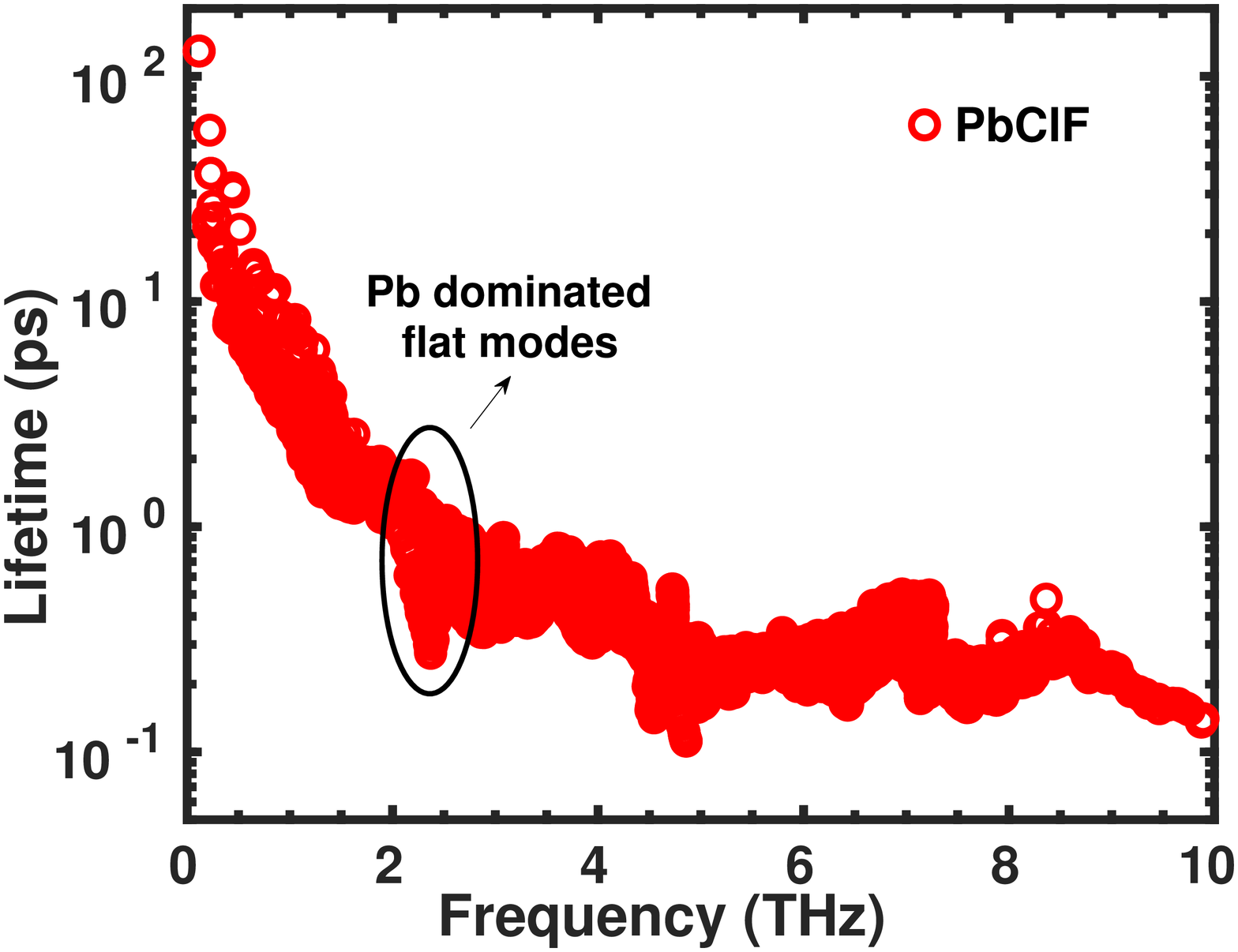}} \\
\subfigure[]{\includegraphics[width=0.5\textwidth]{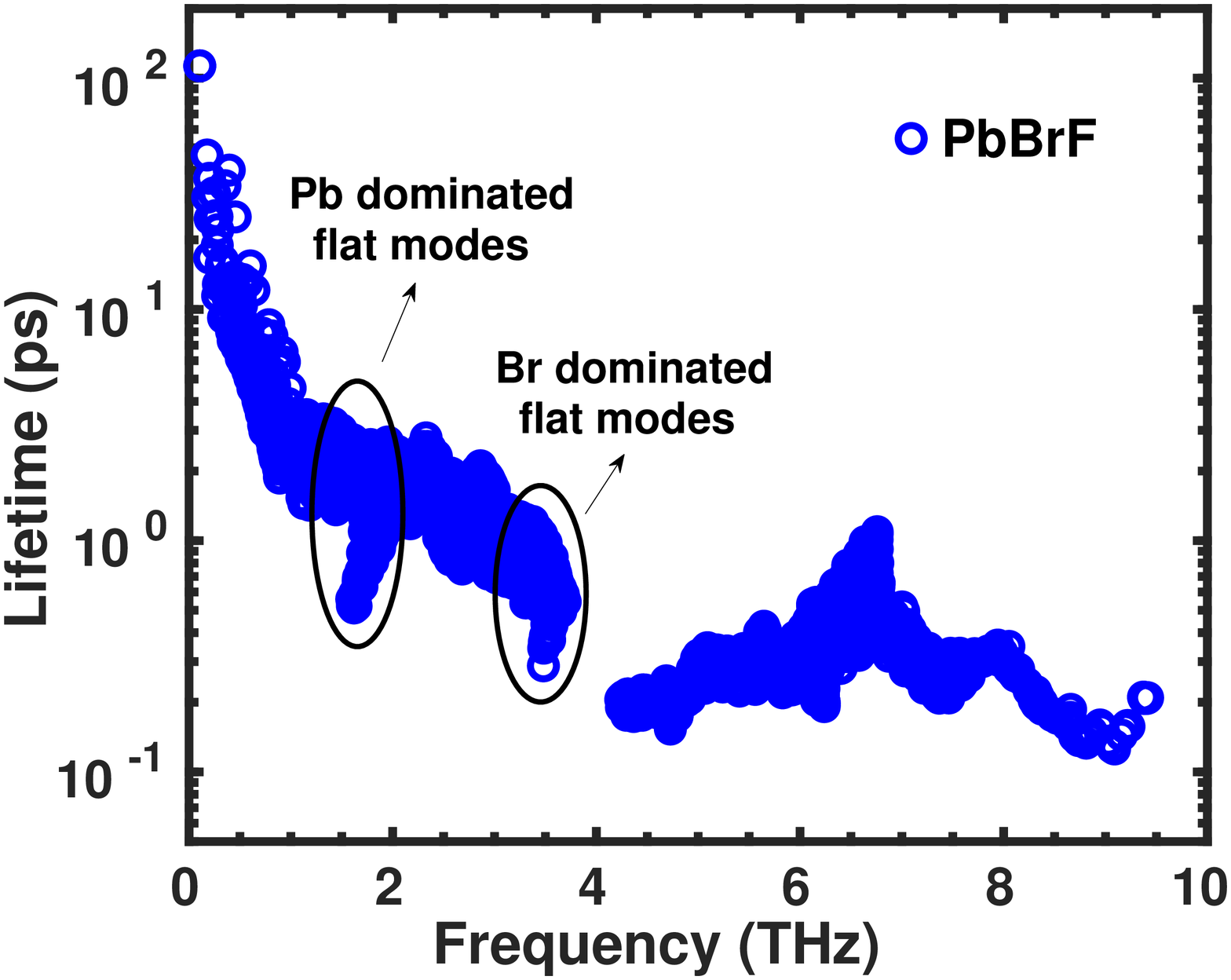}} \\ 
\subfigure[]{\includegraphics[width=0.5\textwidth]{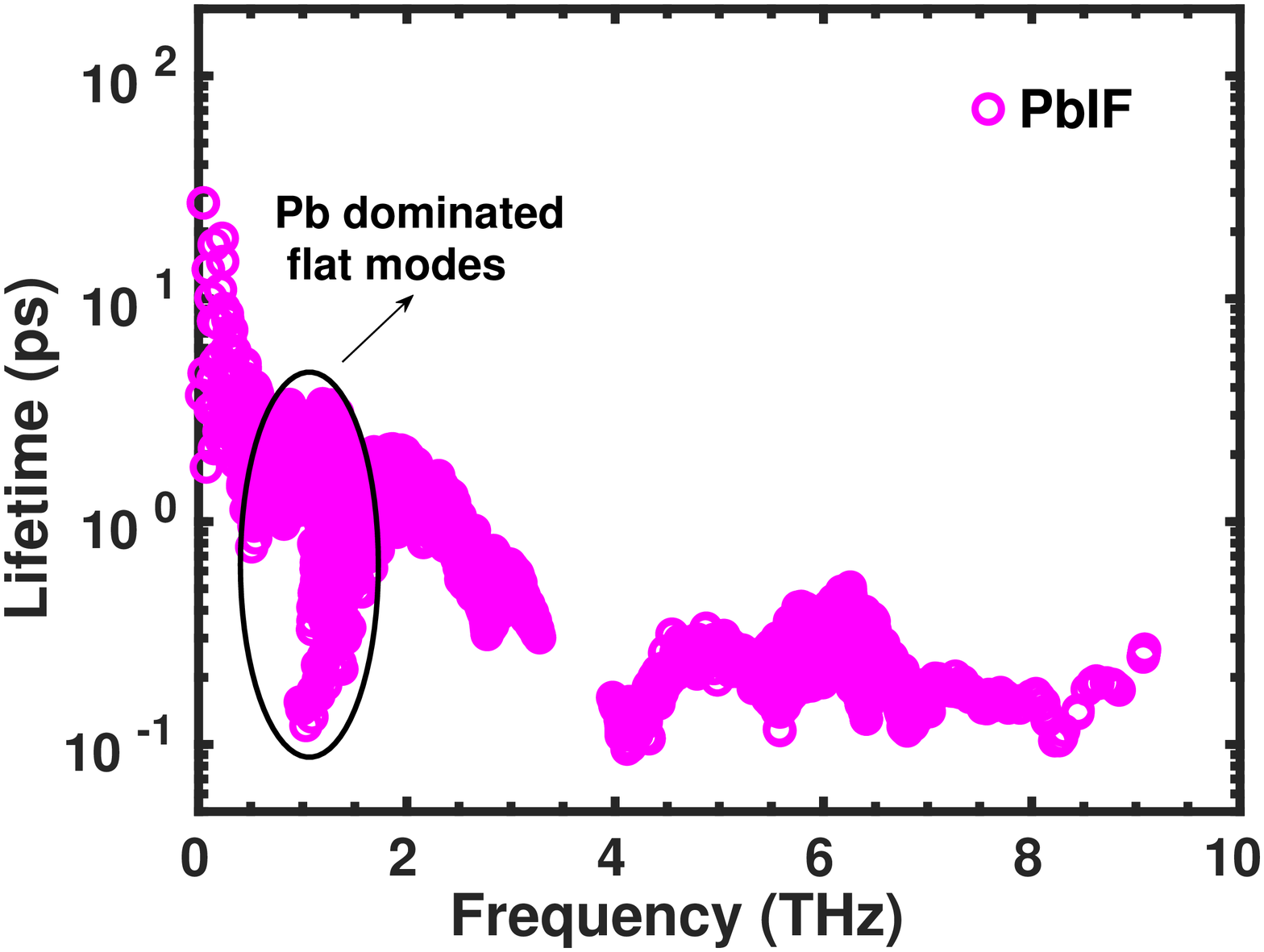}}
\caption{Phonon lifetimes of (a) PbClF, (b) PbBrF, and (c) PbIF as functions of frequency.}
\label{fig:LT}
\end{figure}

\begin{figure}
\centering
\includegraphics[width=0.5\textwidth]{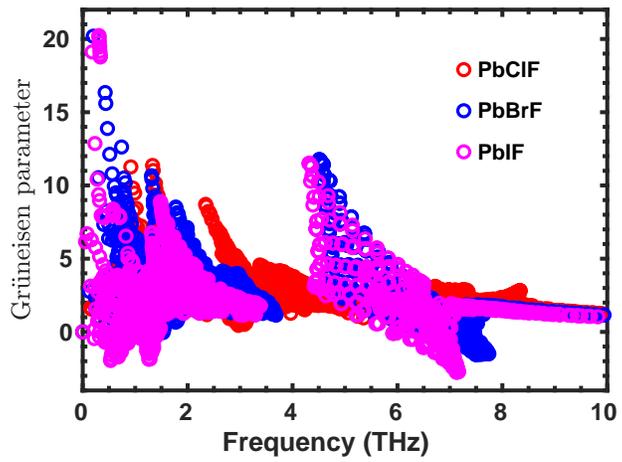}
\caption{Gr\"uneisen parameter of PbXF (X = Cl, Br, I) as a function of frequency.}
\label{fig:Gamma}
\end{figure}

We study the phonon group velocities, lifetimes, and mean free path as functions of frequency at 300 K. For a solid material we have 
\begin{equation}
\kappa_l = \frac{1}{3}{\Huge \int} C(\omega)v_g^2(\omega)\tau(\omega)d\omega,
\end{equation}
where $C(\omega)$ is the spectral volumetric specific heat, $v_g(\omega)$ is the phonon group velocity, and $\tau(\omega)$ is the phonon lifetime. The calculated phonon group velocities (Table \ref{PhGV}) are smaller in the $c$-direction ($\Gamma$-Z) than in the $a$-direction ($\Gamma$-X), as the weak inter-layer vdW bonding prevents phonon transport in the $c$-direction, and they are lower for PbIF than for PbClF and PbBrF (Figure \ref{fig:GV}). We find that the flat phonon bands dominated by vibrations of the Pb and Br/I atoms are associated with relatively low phonon lifetimes (Figure \ref{fig:LT}), which is due to enhanced three-phonon scattering.\cite{BiCuSeO2017} PbIF shows shorter lifetimes of the acoustic and low lying optical phonons than PbBrF and PbClF, hence realizing the lowest $\kappa_l$ values.
The arharmonicity due to the Pb$^{2+}$ 6s$^2$ lone pair can be quantified by the dimensionless Gr\"uneisen parameter ($\gamma$). High $\gamma$ values of both the acoustic and low lying optical phonons represent a strongly anharmonic behavior (Figure \ref{fig:Gamma}). The calculated average $\gamma$ values are 2.67, 2.84, and 2.50 for PbClF, PbBrF, and PbIF, respectively, being higher than in anharmonic lead chalcogenides such as PbS (2.46), PbSe (2.66), and PbTe (1.49).\cite{Xiao2016} 

\begin{figure}
\centering
\subfigure[]{\includegraphics[width=0.5\textwidth]{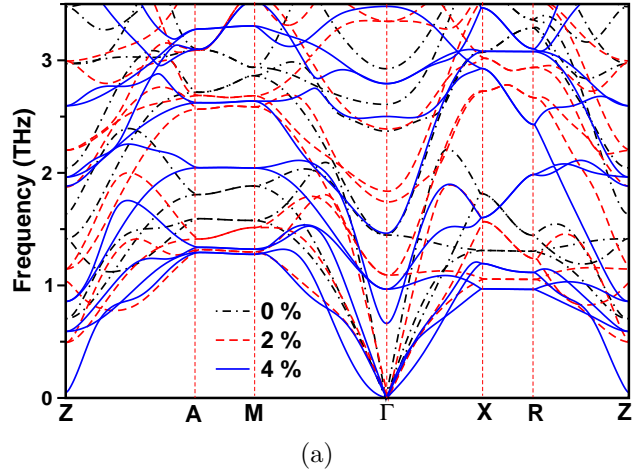}}\vspace{0.5in} \\
\subfigure[]{\includegraphics[width=0.5\textwidth]{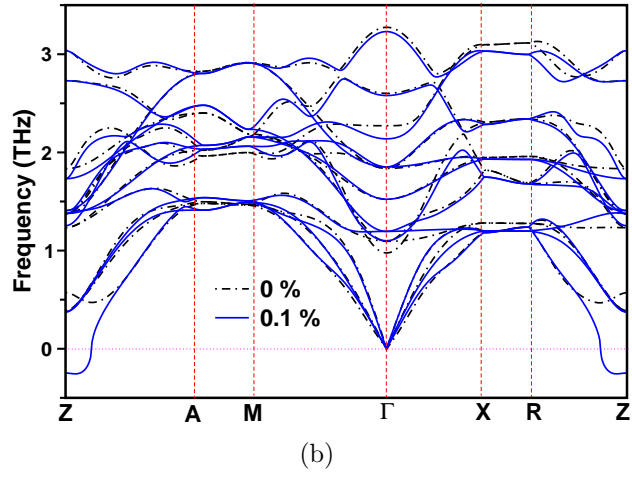}} 
\caption{Phonon dispersion under volume expansion for (a) PbClF (0-4$\%$) and (b) PbIF (0-0.1$\%$) at 300 K. Soft phonon modes are observed in the Z-A and R-Z directions. A lattice instability is predicted for PbIF close to the equilibrium volume.}
\label{fig:FI}
\end{figure}

The obtained high BECs (Table \ref{born}) due to the cross-band-gap hybridization between the Pb $p$-states and X $p$-states suggest that the materials are close to lattice instability. We therefore calculate the phonon dispersions for small (equilibrium) volume expansions from 0.1 to 4$\%$. PbClF, PbBrF, and PbIF become dynamically instable at volume expansions of 4$\%$, 2$\%$, and 0.1$\%$, respectively (Figure \ref{fig:FI}). The fact that PbIF becomes dynamically instable already for a small volume expansion points to proximity to immediate lattice instability, thus resulting in ultralow $\kappa_l$.

\begin{figure}
\centering
\subfigure[]{\includegraphics[width=0.5\textwidth]{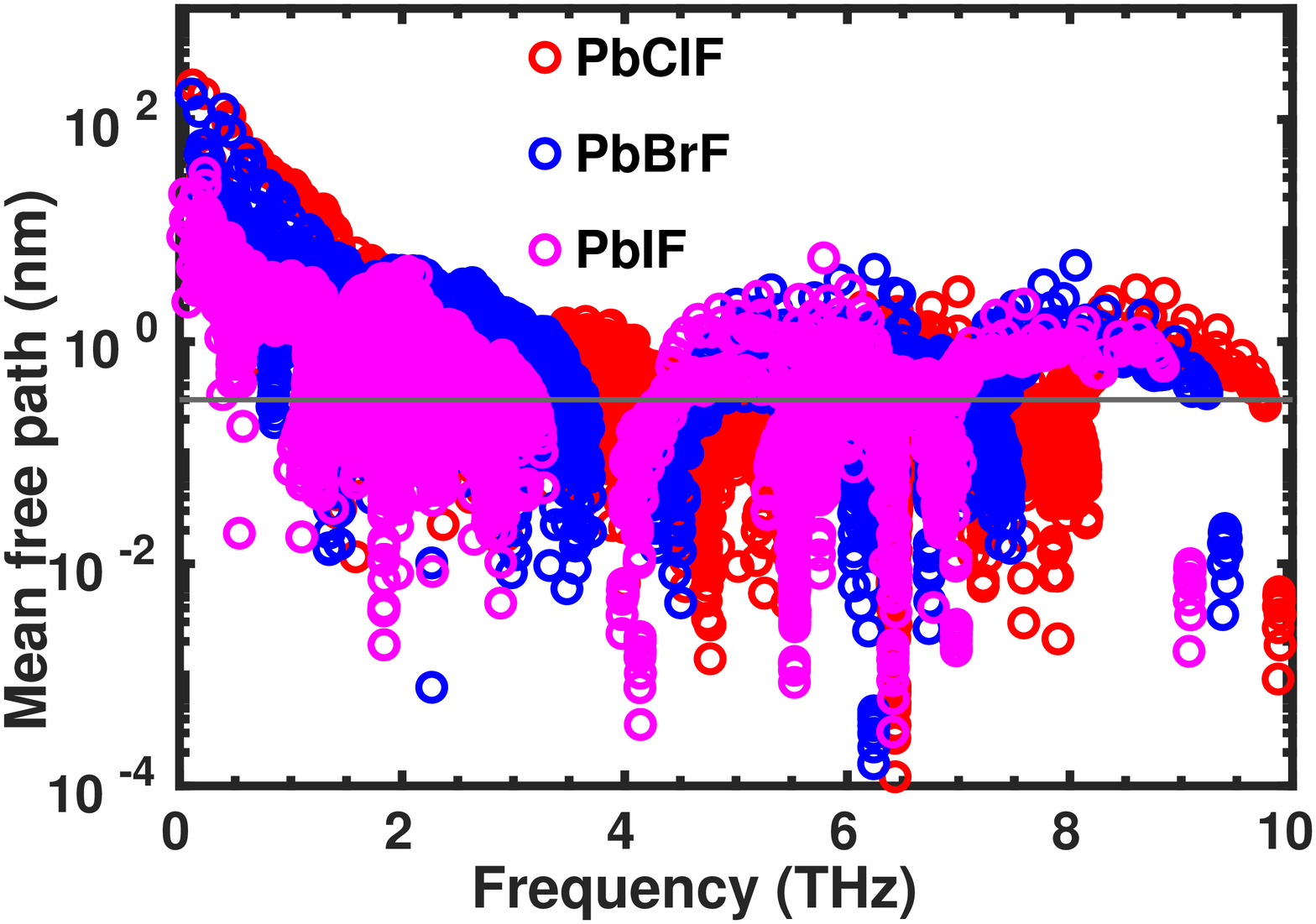}}\vspace{0.3in}
\subfigure[]{\includegraphics[width=0.5\textwidth]{Figures/Fig10b.eps}}
\caption{(a) Phonon mean free path as a function of frequency and (b) temperature dependence of the lattice thermal conductivity using the two channel transport model for PbXF (X = Cl, Br, I).}
\label{fig:MFP}
\end{figure}

Considerable portions of the calculated phonon mean free paths are found to fall below the Ioffe-Regel limit (Figure \ref{fig:MFP}), which implies that these phonons are ill-defined and the BTE, therefore may be insufficient to reproduce the experimental $\kappa_l$ values. To address the presence of ill-defined phonons along with well-defined phonons, we use a phenomenological model that describes the phonon transport in two channels: lattice phonons propagating as wave packets ($\kappa_{BTE}$) and hopping among uncorrelated localized oscillators ($\kappa_{hop}$).\cite{Mukhopadhyay2018} Thus, $\kappa_{l}$ = $\kappa_{BTE}$ + $\kappa_{hop}$. The Cahill-Watson-Pohl model provides
\begin{equation}
\kappa_{hop} = \left(\frac{\pi}{6}\right)^{1/3} k_Bn^{2/3}\sum_{i} v_i\left(\frac{T}{\theta_i}\right)^2\int_{0}^{\theta_i/T} \frac{x^3e^x}{(e^x-1)^2} dx,
\end{equation}
where $v_i$ and $\theta_i$ are the phonon velocity, and Debye temperature of the acoustic branch $i$, respectively (calculated using eq $\ref{debye}$). The two channel transport model reproduces the experimental $\kappa_l$ values for various ultralow $\kappa_l$ materials such as Tl$_3$VSe$_4$,\cite{Mukhopadhyay2018} Tl$_3$VS$_4$,\cite{Mukhopadhyay2019} Tl$_3$TaS$_4$,\cite{Mukhopadhyay2019} TlInTe$_2$,\cite{Wu2019} and TlAgTe\cite{Aamir2020}. At 300 K, the calculated $\kappa_l$ values are 1.66, 1.14, and 0.90 Wm$^{-1}$K$^{-1}$ for PbClF, PbBrF, and PbIF, respectively, exceeding those obtained using ShengBTE by iteratively solving the BTE by 20-50$\%$ (Figures \ref{fig:TC} $\&$ \ref{fig:MFP}b), which demonstrates that the two channel model is essential to calculate $\kappa_l$ accurately for materials with ultralow $\kappa_l$, consistent with previous reports.\cite{Mukhopadhyay2018,Mukhopadhyay2019,Wu2019,Aamir2020} For instance, the $\kappa_l$ value calculated by the BTE is 0.158\cite{Mukhopadhyay2018} (0.17\cite{Aamir2020}) Wm$^{-1}$K$^{-1}$ and that calculated by the two channel model is 0.368\cite{Mukhopadhyay2018} (0.43\cite{Aamir2020}) Wm$^{-1}$K$^{-1}$ for Tl$_3$VSe$_4$ (TlAgTe),\cite{Mukhopadhyay2018,Kurosaki2007} being comparable with the experimental value of 0.3 $\pm$ 0.05 (0.44) Wm$^{-1}$K$^{-1}$. 





\section{Conclusions}
We have systematically investigated the chemical bonding, elastic properties, electronic properties, lattice dynamics, and phonon transport of PbXF (X = Cl, Br, I)  using a combination of density functional theory, Boltzmann transport theory, and a two channel transport model. The obtained low elastic moduli (B, G, and E) and high compressibility suggest that the materials are comparably soft, especially PbIF. Low sound velocities and Debye temperatures are indicative of low $\kappa_l$ values. Enhanced BECs due to cross-band-gap hybridization bring the materials close to lattice instability. In particular, PbIF becomes instable at 0.1$\%$ volume expansion. 
The computed phonon frequencies show a red-shift for increasing size of the X anion and a large mass difference between F and Pb/X opens up a phonon band gap in PbBrF and PbIF. The layered crystal structure (bonding heterogeneity) results in low phonon group velocities and low phonon transport in the out-of-plane direction. Large $\gamma$ values ($\textgreater$ 2.5) are indicative of a strong anharmonicity. Low dispersions of the low lying optical phonons result in enhanced phonon scattering, which causes a reduction of the phonon lifetimes, especially in PbIF. The $\kappa_l$ values obtained by solving the BTE using ShengBTE are 1.31, 0.78, and 0.61 Wm$^{-1}$K$^{-1}$ at 300 K for PbClF, PbBrF, and PbIF, respectively, while those obtained by the two channel model are 1.66, 1.14, and 0.90 Wm$^{-1}$K$^{-1}$ (20-50$\%$ higher). Overall, the interplay between bonding heterogeneity, lone pair induced anharmonicity, and the high mass differences between F and Pb/I results in softening of the acoustic and low lying optical phonons and causes the ultralow $\kappa_l$ of PbIF. Our study demonstrates that a material should possess bonding heterogeneity, one or more lone pair containing cations, and high mass differences to achieve ultralow $\kappa_l$ values, which is critically important to design functional materials for future thermal energy applications.

\section{Supporting Information}
Eigenvectors of the acoustic and optical phonon modes of PbIF (Figures S1 $\&$ S2) and ground state structural properties (Tables S1 $\&$ S2). (PDF)

\section{Acknowledgments}
N.Y. thanks the Stony Brook Research Computing and Cyberinfrastructure as well as the Institute for Advanced Computational Science at Stony Brook University for access to the high-performance SeaWulf computing system, which was made possible by a  National Science Foundation grant (1531492). The research reported in this publication was supported by funding from King Abdullah University of Science and Technology (KAUST). S.C.R.R. thanks Rajiv Gandhi University of Knowledge Technologies (RGUKT) Basar for providing computational facilities.

\bibliography{Refs.bib}

\clearpage
{}

{}


{}

{}


{}

{}

\clearpage


{}

\clearpage

\begin{table}[tbp]
\caption{Bader charges of the constituent elements of PbXF (X = Cl, Br, I).}
\label{ELFT}
\begin{tabular}{ccccc} \hline
Element   &  Formal charge  &   PbClF     &  PbBrF    &   PbIF      \\ \hline
Pb        &   +2            &   +1.4    &  +1.4  &   +1.3     \\
F         &   -1            &  -0.8     &  -0.8   &  -0.8     \\
Cl/Br/I   &   -1            &  -0.7     &  -0.6   &  -0.5     \\  \hline
\end{tabular}
\newline
\end{table}


\begin{table}[tbp]
\caption{Second order elastic constants (in GPa) of PbXF (X = Cl, Br, I).}
\label{elastic}
\begin{tabular}{cccccccc} \hline
&        &  C$_{11}$  & C$_{33}$   &  C$_{44}$    &  C$_{66}$   &  C$_{12}$  & C$_{13}$ \\ \hline
PbClF    & This work$^a$    &  88      &    71    &     21       &   31      &   46     &  46    \\
         & This work$^b$    &  80      &    56    &     18       &   26      &   40     &  39     \\ 
         &Others$^c$  &  89     &    75    &     28       &   31      &   29     & 41   \\ 
PbBrF    & This work$^a$    &  80      &    47    &     21       &   32      &   44     &  39    \\
         &  This work$^b$   &  74      &    36    &     19       &   26      &   39     &  34    \\   
         &Others$^c$  & 80    &   68   &     25       &   28      &   26    & 37  \\ 
PbIF     & This work$^a$    &  60     &    15    &     10       &   23      &   22     &  15    \\
         & This work$^b$    &   59     &    13    &     12       & 23        &   23     &  16    \\         
         &Others$^c$  &  67     &   56   &    21       &  23     &   21     & 30  \\ \hline
\end{tabular}
\newline
$^a$At the PBEsol equilibrium volume \\
$^b$At the experimental volume \\
$^c$Ref~\citen{SIESKIND199875}.
\end{table}

\begin{table}[tbp]
\caption{Bulk modulus (B$_V$, B$_R$, B$_{VRH}$ in GPa), shear modulus (G$_V$, G$_R$, G$_{VRH}$ in GPa), Young's modulus (E in GPa), density ($\rho$ in gr/cc),  sound velocity (v$_l$, v$_t$, v$_m$ in km/s), and Debye temperature ($\theta$ in K) of PbXF (X = Cl, Br, I).}
\label{ployc}
\begin{tabular}{ccccc} \hline
  &   PbClF     &  PbBrF    &   PbIF      \\ \hline
B$_V$        &    58.1    &  50.2    &   26.4   \\
B$_R$        &    57.9    &  45.6    &   14.3  \\
B$_{VRH}$    &    57.9    &   47.9   &   20.3   \\ 
G$_V$        &    21.8    &   20.4    &   14.4   \\ 
G$_R$        &    20.8    &  17.3    &  9.9                    \\
G$_{VRH}$    &    21.3    &  18.8    &   12.2  \\
E          &    56.9    &  49.9    &  30.4                    \\
$\rho$     &   7.3      &  7.8    & 7.5                    \\
$v_l$      &   3.4      &  3.1     &   2.2  \\    
$v_t$      &    1.7      &  1.6     &   1.3   \\
$v_m$      &    1.9      &   1.7     &   1.4   \\ 
$\theta$ &    210.7      &  185.7     &   142.1   \\ \hline
\end{tabular}
\newline
\end{table}

\begin{table}[tbp]
\caption{Born effective charge (Z*), electronic dielectric constant ($\epsilon^{\infty}$), and ionic  dielectric constant ($\epsilon^0$) of PbXF (X = Cl, Br, I).  The symbols $\parallel$ and $\perp$ represent the out-of-plane and in-plane directions, respectively.}
\label{born}
\begin{tabular}{cccccc} \hline
         &       Atom      &     Z*$_\parallel$ &  Z*$_\perp$    &  $\epsilon_\parallel^{\infty}$ ($\epsilon_\perp^{\infty}$)  & $\epsilon_\parallel^0$ ($\epsilon_\perp^0$)       \\ \hline    
PbClF       &       Pb        &     3.42           &     3.74         &   4.41 (5.00)  &    20.95 (38.36)\\
            &       Cl        &     -1.37          &    -1.99         &                     \\
            &       F         &     -2.05          &    -1.76         &                     \\
PbBrF       &       Pb        &     3.45           &     3.98         &   4.86 (5.88)  &    19.87 (42.57)\\
            &       Br        &     -1.22          &    -2.10        &                     \\
            &       F         &     -2.22          &    -1.88         &                     \\
PbIF        &       Pb        &     2.91          &     4.20         &   4.71 (6.73)  &   5.00 (41.37) \\
            &       I         &     -0.76           &    -2.17         &                      \\
            &       F         &     -2.15          &    -2.03         &                      \\ \hline
\end{tabular}
\newline
\end{table}

\begin{table}[tbp]
\caption{Electronic band gaps (in eV) with and without SOC of PbXF (X = Cl, Br, I) using the PBE functional, HSE functional, and TB-mBJ potential compared with reflection spectrum measurements\cite{Liu2005} and calculations.}
\label{bandgaps}
\begin{tabular}{cccccc} \hline
            &  Method          &   PbClF     &  PbBrF   &   PbIF      \\ \hline
This work   & PBE              &    3.5      &  2.6     &   1.8  \\    
            & PBE+SOC           &    3.2      &  2.3     &   1.4   \\
            & TB-mBJ           &    4.8      &  3.9     &   2.8    \\ 
            & TB-mBJ+SOC        &    4.7      &  3.8     &   2.6   \\ 
            & HSE              &    4.8      &  4.0     &   3.6   \\
            & HSE+SOC         &    4.7      &  3.9     &   3.4    \\
Others       & LDA$^a$          &    3.23     &  2.39     &   1.73  \\ 
            & PBE$^{a, b}$ &    3.49, 3.5  & 2.63    &   2.0   \\                 
            & EV$^{a, c}$  &    4.28, 4.32  &  3.31, 3.37  & 2.61, 2.48   \\ 
            & GW$^b$           &    5.0      &  --     &   --  \\
     Experiment$^b$        &        &    5.2     &  --     &   --  \\  \hline
\end{tabular}
\newline
$^a$Ref~\citen{hajHassan2004}.
$^b$Ref~\citen{Liu2005}.
$^c$Ref.~\citen{Reshak2007}. 
\end{table}

\begin{table}[tbp]
\caption{Phonon group velocities (in km/s) of the acoustic branches of PbXF  (X = Cl, Br, I) in the $\Gamma$-X and $\Gamma$-Z directions.}
\label{PhGV}
\begin{tabular}{ccccc} \hline
   &   &  PbClF  &  PbBr  &  PbIF  \\ \hline
$\Gamma$-X       &  TA1       &  1.56   &  1.59  &  1.13  \\
                 &  TA2       &  1.95   &  1.92  &  1.80  \\
                 &  LA        &  3.50   &  3.12  &  2.37  \\
 $\Gamma$-Z       &  TA1       &  1.60   &  1.52  &  0.78  \\
                 &  TA2       &  1.60   &  1.52  &  0.78  \\
                 &  LA        &  2.99   &  2.55  &  1.26  \\ \hline
\end{tabular}
\newline
\end{table}

\clearpage
\begin{center}
{\bf \Large TOC Graphic}
\end{center}

\begin{figure}
    \centering
    \includegraphics[width=6.5in,height=4.5in]{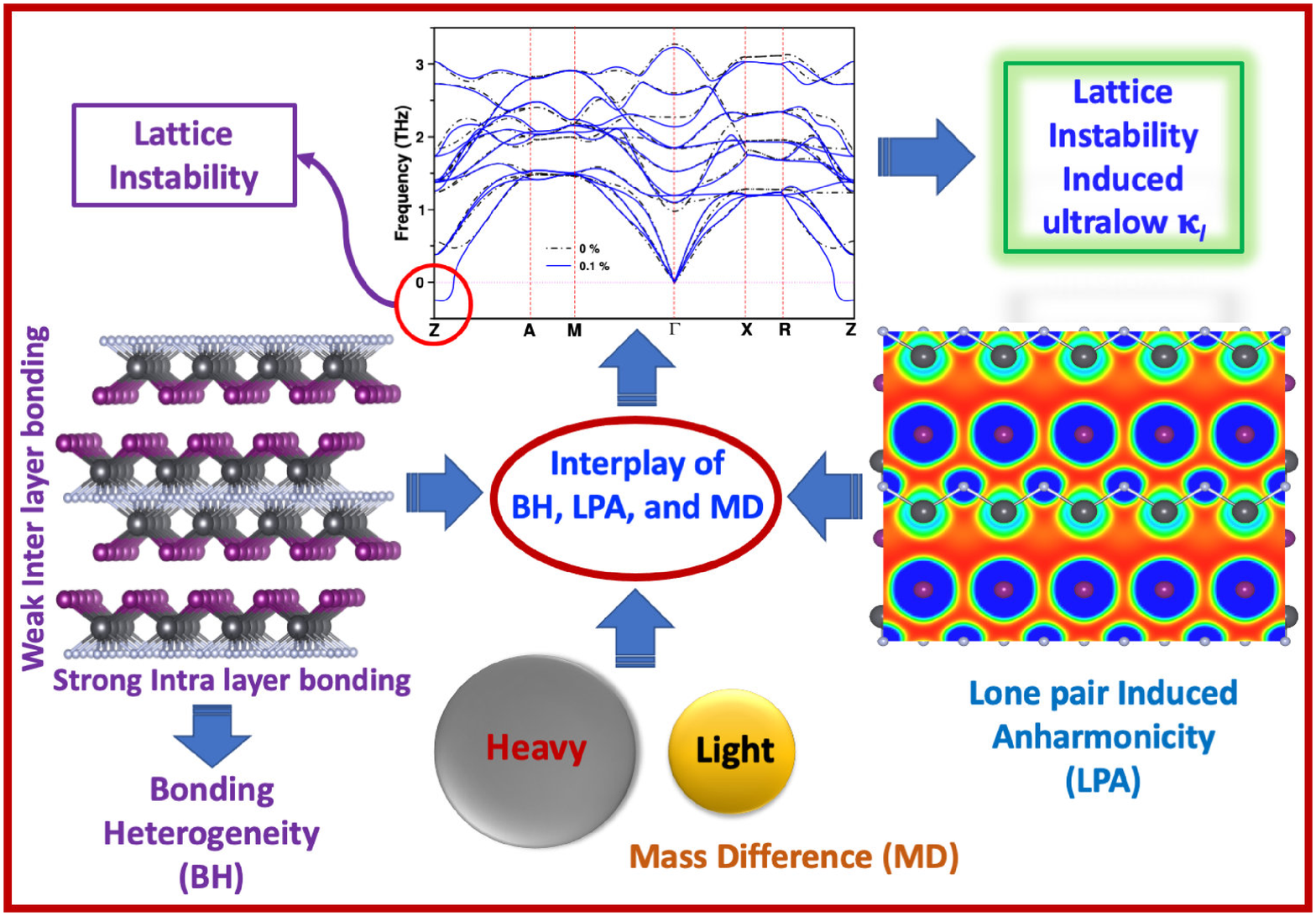}
    \caption*{Interplay of bonding heterogeneity, lone-pair induced anharmonicity
and constituent elements with high mass difference within the same crystal structure
aid in designing ultralow $\kappa_l$ materials for thermal energy management applications.}
    \label{fig:TOC}
\end{figure}

\end{document}




\clearpage

\begin{figure}
\centering
\includegraphics[width=6.0in,height=8.5in]{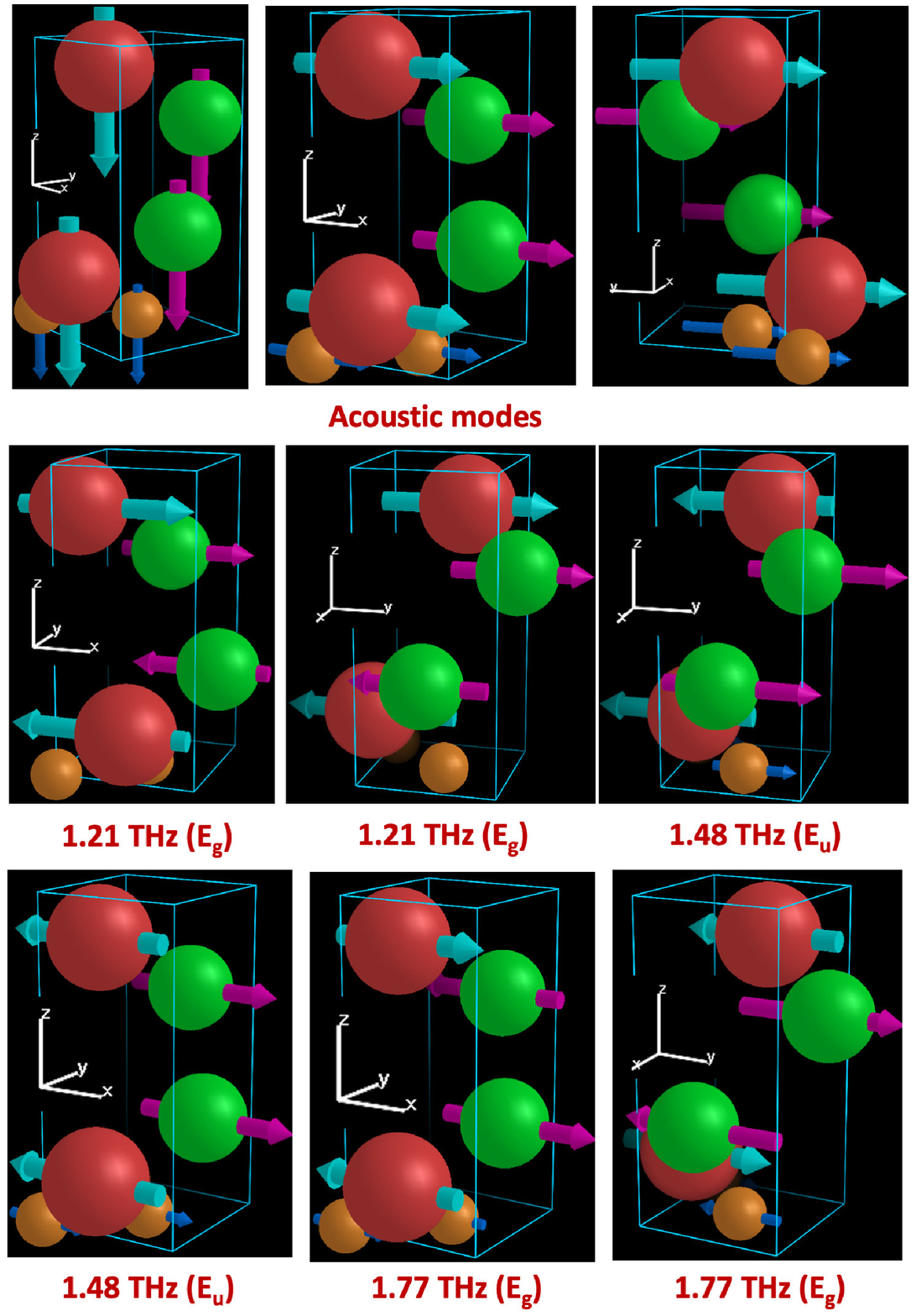}
\caption{Eigenvectors of the acoustic and low frequency optical phonons of PbIF. E$_g$ and E$_u$ are doubly degenerate.}
\label{fig:TC}
\end{figure}{}
\begin{figure}
\centering
\includegraphics[width=6.0in,height=8.5in]{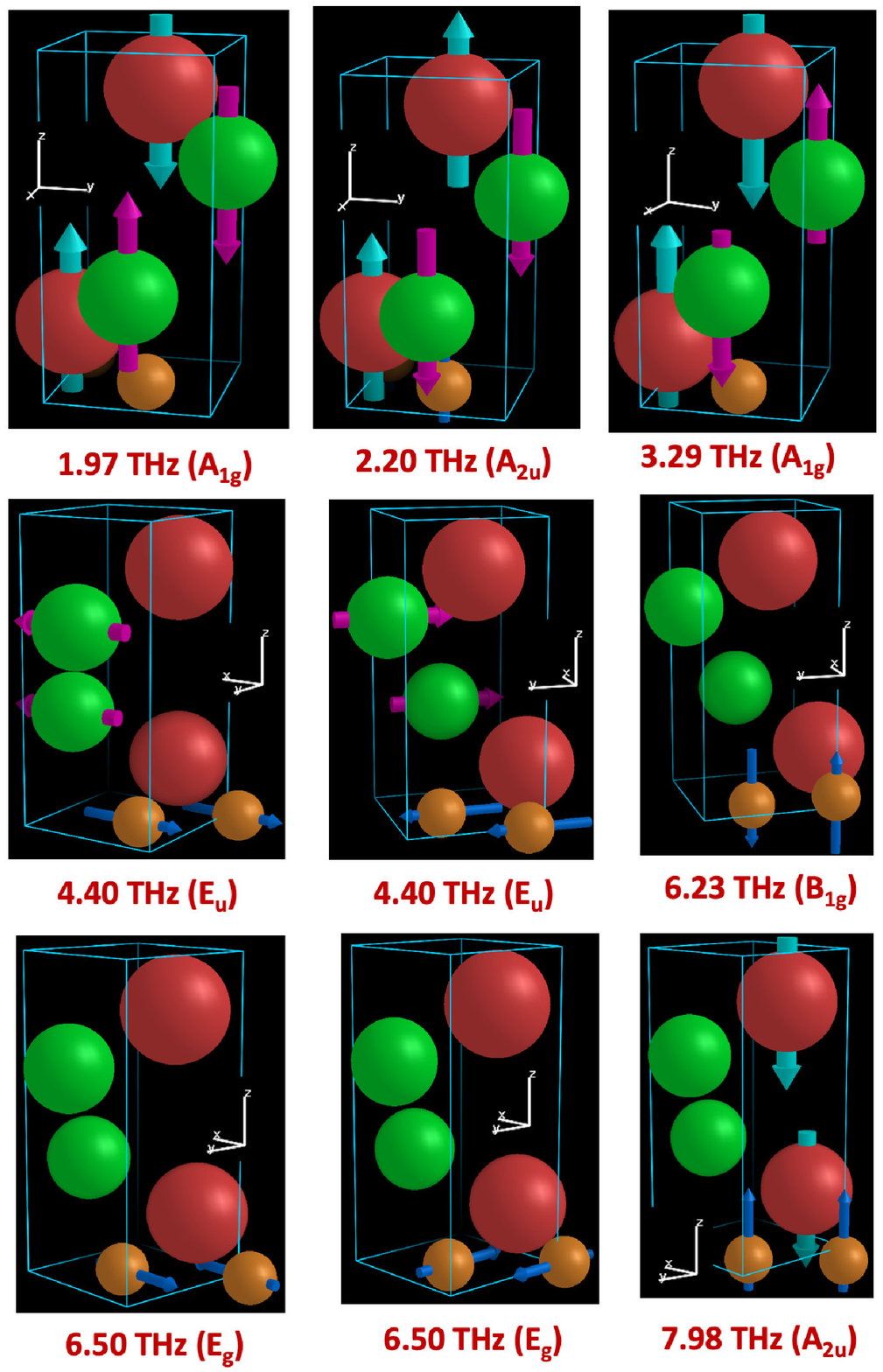}
\caption{Eigenvectors of the intermediate and high frequency optical phonons of PbIF. E$_g$ and E$_u$ are doubly degenerate.}
\label{fig:TC}
\end{figure}{}



\clearpage
\begin{table}[tbp]
\caption{Ground state lattice parameters (a, c in \AA) and volume (V in \AA$^3$) of PbIF using standard and dispersion corrected DFT methods.}
\label{table1}
\begin{tabular}{ccccc} \hline
Material &  Method       &   a      &    c    &  V  \\ \hline
PbIF     &  PBEsol         &  4.161   &  9.058  &  156.83 \\
         &  DFT-D2         &  4.248   &  8.580  &  105.33 \\
         &  optB88-vdW     &  4.233   &  8.976  & 160.83 \\ 
         &  optB86b-vdW    &  4.218   &  8.888  & 158.13 \\
         &  Experiment$^a$      &  4.23   &  8.77  &  156.92  \\   \hline  
\end{tabular}
\newline
$^a$Ref~\citen{Decremps1999}. 
\end{table}


\begin{table}[tbp]
\caption{Ground state structural properties of PbXF (X = Cl, Br, I) using the PBEsol functional compared with X-ray diffraction data and other calculations.}
\label{table1}\scalebox{.85}{
\begin{tabular}{ccccccccccc} \hline
Material  & Method     &   a     &   c     &    V      &   $v$      &  $u$    \\ \hline
PbClF     & This work  & 4.066   & 7.186   & 118.80    &  0.2077  & 0.6468  \\
          & Experiment$^a$  & 4.11    & 7.246   & 122.4     &  0.2058  &  0.6497  \\
          & Experiment$^b$  & 4.1077  & 7.2282  & 121.968   &  -  &   -     \\
          & Others     & 4.163$^c$, 4.404$^d$  & 7.340$^c$, 7.079$^d$ & -  & 0.2075$^c$, 0.191$^d$    & 0.6463$^c$,  0.66$^d$   \\
          &            & 4.151$^e$ &  7.354$^e$  &  -    &  0.2075$^e$ &  0.6463$^e$  \\
PbBrF     & This work  & 4.146    & 7.541   & 129.62   & 0.194  & 0.647  \\
          & Experiment$^a$      & 4.18    & 7.59    & 132.62         & 0.195  & 0.650  \\
          & Others     & 4.231$^c$, 4.451$^d$   & 7.760$^c$, 7.202$^d$ &  -  & 0.1945$^c$, 0.187$^d$  & 0.6469$^c$,  0.664$^d$,   \\
          &            & 4.232$^e$              &  7.70$^e$         &    -   &  0.1945$^e$              &  0.6469$^e$ \\
PbIF      & This work  & 4.161    &  9.058   & 156.83   & 0.161  & 0.666  \\ 
          & Experiment$^f$ & 4.23     &  8.77   &  156.92  & 0.167  & 0.65 \\
          & Others     & 4.250$^c$, 4.248$^e$ & 9.152$^c$, 8.87$^e$  &  & 0.1666$^c$, 0.1665$^e$ & 0.6578$^c$, 0.6577$^e$  \\ \hline
\end{tabular}
}
\newline
 $^a$Ref~\citen{Pasero1996}. $^b$Ref~\citen{2D-CNR-Lat-Param}. $^c$Ref~\citen{hajHassan2004}. $^d$Ref~\citen{Mittal-LAt-Parameters}.  $^e$Ref~\citen{Reshak2007}. $^f$Ref~\citen{Decremps1999}.
\end{table}

\clearpage
\bibliography{Refs.bib}